# Microfluidic Bioelectrical Impedance Drug Delivery Device for Patients with Acute Exacerbations of Chronic Obstructive Pulmonary Disease

**Evan Carroll[1], Nicholas Recchione[1], Jean Paul Rojas Henao[1], Vincent G. Capone[1]**
**Sleiman R. Ghorayeb, Ph.D.[1, 2, 3]**

[1] Department of Bioengineering, Fred DeMatteis School of Engineering and Applied Science, Hofstra University, Hempstead, NY 11549, USA
[2] Ultrasound Research Laboratory, Hofstra University, Hempstead, NY 11549, USA
[3] Departments of Radiology and Molecular Medicine, Donald and Barbara Zucker School of Medicine at Hofstra/Northwell, Hofstra University, Hempstead, NY 11549, USA

*Abstract*—Inhalers with corticosteroids and muscle relaxants are prescribed by pulmonologists to mitigate bronchospasms that happen due to tightening of the chest. An adhesive patch attached to a patient with chronic obstructive pulmonary disease can relieve episodes of bronchospasms in the event the prescribed inhaler is not nearby. This paper shows the design and programming created to manufacture such a device. *Methods:* Electrical components were connected to a 3D model created in AutoCAD Fusion 360 in the order of inches. The model was imported, and simulation tests were evaluated in COMSOL Multiphysics. The device was 3D printed in the Makerspace at Hofstra University and tested for functionality. *Results:* The proposed design is a macromodel prototype of the micromodel adhesive patch that will attach above the sternocleidomastoid (SCM) muscle on one side of the neck and be created in a microelectronics laboratory. *Conclusion:* The designed bioelectrical impedance device can detect severe muscle contractions related to the tensing and thickening of the SCM to deliver muscle relaxants to relieve acute exacerbations in patients with COPD. *Significance:* With this microfluidic drug delivery device even if the patient has forgotten to carry their inhaler, they are able to be relieved of the suffocating bronchospasm episode quickly and inhale their prescribed medication.

## I. INTRODUCTION

Chronic obstructive pulmonary disease (COPD) is a chronic inflammatory lung disease where irreversible damage is done to the lungs from other conditions that block airflow and make it difficult to breathe such as bronchitis or emphysema. Acute exacerbations because of COPD are triggered mainly by respiratory viruses and bacteria which infect the lower airway and increase inflammation. Although COPD and asthma are incurable diseases, pulmonologists and primary care physicians commonly prescribe combination inhalers to mitigate and relieve complications from shortness of breath and bronchospasm attacks. However, other forms of therapeutic palliative care have not been explored in depth. With the increasing number of people acquiring these chronic respiratory diseases, it is important to pay more attention to the needs of patients who grow more resistant to the positive effects of inhalers alone. This proposed solution is to design a wearable device that adheres to the side of the neck above the sternocleidomastoid (SCM) region to administer muscle relaxant medication and relieve pain and stress faster than the traditional effects of prescription inhalers.

Asthma and COPD are heterogeneous diseases that are identified by abnormal airway functions with variable expiratory airflow limitations [1]. They overlap in more ways than one which makes it more feasible to use similar medication in both cases. Eosinophils, mast cells, neutrophils, macrophages, CD8+ cells, and many antibodies including IgE behave similarly when patients are diagnosed with asthma and/or COPD. Although one can be genetically susceptible to developing these diseases or inherit early childhood lung disorder, many patients have a higher chance of acquiring COPD and asthma as a result of chronic exposure to cigarette smoke, air pollutants, and biomass fuels. Vascular alterations from asthma lead to prominent features such as dilatation of blood vessels in bronchial mucosa, congestion, and edema in the airway wall. With COPD, pulmonary muscular arterial intimal thickness increases due to the smooth muscle cells that proliferate longitudinally and lack differences between cellular and extracellular matrix components, i.e. elastin, collagen, and proteoglycans. The medial muscles increase in thickness which leads to a reduced arterial luminal diameter [1].

Lung hyperinflation causes COPD patients to have difficulty inhaling as the diaphragm becomes flattened and weakened. Increased accessory respiratory muscles, especially from the SCM, compensate for the weakened respiratory function [2]. Ultrasonography (US) is a noninvasive technique whose reliability for the evaluation of changes in muscle thickness has been confirmed. Shiraishi et al. [2] obtained clinical data associating SCM thickening fraction (SCM TF), which they calculated using US, and exercise tolerance in COPD patients. They aimed to evaluate the end-expiration SCM thickness and the end-expiration to resting inspiration and end-expiration to end-inspiration SCM thicknesses using US between COPD patients and control subjects. Their data was then interpolated into a graph that was used to determine the threshold thickness in the SCM muscle of a COPD patient undergoing acute exacerbations.

The American Lung Association lists different types of medicine that are commonly prescribed to treat COPD: bronchodilators, anti-inflammatory, combination medicines, antibiotics, vaccinations. Long-term patients with COPD have experienced the best results from combination medicine treatments which are generally covered by insurance and are easily accessible. In many patients, healthcare providers recommend combination therapy where two different medications are combined in one inhaler or nebulizer treatment. Combination medications include inhaled corticosteroids (ICS) and long-acting beta-2-agonist (LABA), long-acting muscarinic antagonist (LAMA) and LABA, or short-acting beta-2-agonists (SABA) and short-acting muscarinic antagonists (SAMA). In cases with severe airflow limitation with more symptoms and blood eosinophil counts greater than or equal to 100 cells per microliter, healthcare providers recommend triple therapy combining ICS, LABA, and LAMA into one medication [3].

After extensive research, the drug type chosen to be delivered through the device would be a muscle relaxant combination of LABA and LAMA medications. They were ranked the most effective to reduce moderate to severe and severe exacerbations compared to the other groups [4]. The study results found were not applicable to those with a milder form of COPD since they do not usually require maintenance inhaler therapy.

Most biological matter is conductive or makes use of electricity, so changes in the conductivity of human tissue such as the skin and muscle can be recorded [5]. The design of the Bioelectrical Impedance Drug Delivery Device takes advantage of this. As the SCM muscle thickness changes due to Acute Exacerbations from COPD, there will be variations in its electrical resistance/properties that can be detected by an impedance sensor that will be on the device. As the impedance decreases (due to the thinning of the SCM muscle) and reaches a specified threshold, it would be determined that a patient is suffering from acute exacerbations, meaning that the device should be activated. For this to be simulated, an equivalent resistance circuit that can represent the layers of skin (epidermis and dermis) and muscle (for the SCM muscle) had to be created. After determining the resistance and capacitance of each layer and putting them together to represent the patch-neck interface, a simplified circuit/model could be made with values dependent on the age of the human skin and SCM muscle [6]. R_Skin and C_Skin represents the resistance and capacitance, respectively, of the epidermis layer and the dermis layer combined in a circuit (*Fig. 1*), and R_Muscle represents the resistance of the desired muscle layer, in this case, the SCM muscle. Since the resistance of skin differs with age, a patient's doctor may have to set the device accordingly to be effective for the patient.

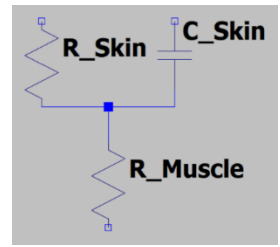

Figure 1. Simplified Circuit to represent equivalent resistance and capacitance of human skin (epidermis and dermis) and muscle [7].

## II. METHODS AND MATERIALS

### A. Modeling and Design

Autodesk Fusion 360 is a computer-aided design (CAD) software that includes design and 3D modeling tools used to blueprint and construct the COPD medical device. Although the goal is to shrink the device to the microscale, a macromodel was designed to hold a 1" diameter 40 kHz transducer, ⅞" outer diameter (OD) duckbill valve, and ¼" OD inlet tube fittings (*Fig. 2*). After many revisions, the model compactly fit within a 2" x 3" rectangular area and was 3D printed. The tube fittings model was found on McMaster-Carr, an American engineering supplies company. The transducer was bought from Calectro GC Electronics and labeled to be used for remote controls and intrusion alarms. The duckbill valve and tubing were bought from Amazon marketplace.

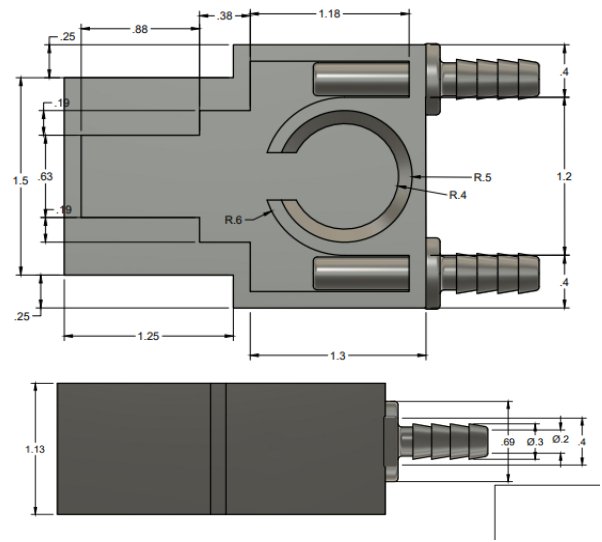

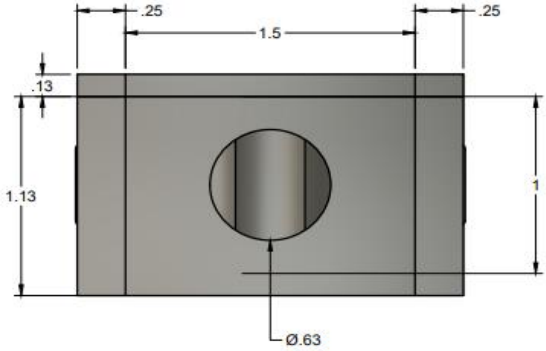

Figure 2. Autodesk Fusion 360 drawing of macromodel (top, side, back views, respectively) with inlet tube fittings with dimensions in inches.

### B. Biosensing Effects of COPD in the SCM

#### 1) SCM Detection and Device Activation

Inspiratory muscle training is used to observe the activation and thresholds of the diaphragm and accessory muscles in COPD patients with inspiratory muscle weakness. Results were obtained to hypothesize and predict at what thickness and intensity the device would need to activate to interpret the patient undergoing acute exacerbations. Fig. 3 shows SCM thickness in patients with COPD and control subjects [2]. Using this information, we were able to create an Excel sheet with the necessary time points needed to simulate an accurate depiction of the respiration of a patient with COPD. LabVIEW was utilized to simulate a biosensing device that employs ultrasound to measure SCM muscle thickness during respiration and compare it against a specific threshold, subsequently sending a signal to another device if the threshold is met.

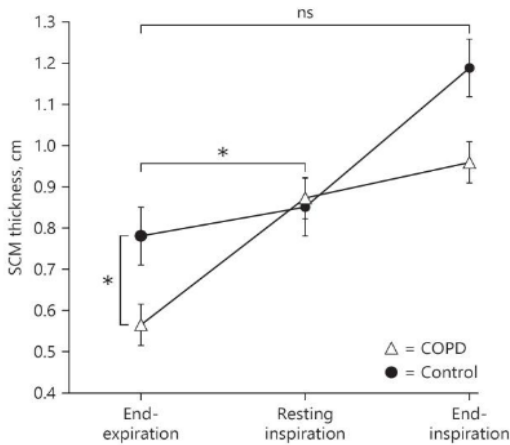

Figure 3. SCM thickness in patients with COPD and control subjects. [2]

#### 2) Raw Data

After the Excel sheet containing the necessary time points was generated, it was converted into a text file. The file was introduced into the LabVIEW program by creating a file path. The data from the file was converted into a 2D array for the front panel and into two separate 1D arrays to be used in the ultrasound echo ranging equation. Fig. 14 shows the front panel with the file path created for the raw data and Fig. 12 shows the block diagram for the raw data array conversion.

#### 3) Analysis

An ultrasonic sensor uses the echo ranging equation to determine the thickness of the SCM based on the time it takes for an ultrasound pulse to travel through the SCM muscle and back. The basic principle relies on the speed of sound in each medium, which in this case was skin and muscle tissue. This equation was replicated in LabVIEW using numeric functions and constants. A For Loop was used to iterate through the 1D arrays containing the time points of the ultrasound pulse. For each iteration the time points are fed into the Subtract Function to determine their difference, yielding the Transit Time measured in microseconds. This transit time was then multiplied by the speed of sound through muscle tissue, which is 1540 m/s, and the result was divided by two to calculate the thickness of the SCM. Finally, the thickness was converted into centimeters for the final measurement and stored into a new array as well as plotted on a graph. Fig. 15 shows the front panel and Fig. 12 shows the block diagram for the analysis.

$$Transit\ Time\ =\ t2 - t1 \qquad (1)$$

$$Tissue\ Thickness\ (cm) = \left(\frac{(Transit\ Time\ \times\ Velocity\ of\ Sound)}{2}\right) \div 10{,}000 \qquad (2)$$

#### 4) LabVIEW Signal Transmission

An Arduino Uno unit was used to receive the signal from LabVIEW. To establish a LabVIEW-Arduino interface another *For Loop* was used to iterate through the thickness values produced by the echo ranging equation. During each iteration, the thickness value was displayed on the front panel and compared to the threshold thickness of 0.6 cm or less. Simultaneously, a serial connection to the Arduino was established using the specified serial communication port on the front panel. Once LabVIEW connected to the device, a case structure was utilized to determine if a signal should be sent to the Arduino based on whether the threshold was met. If the threshold was met, the value was written to the digital output channel on the Arduino, as specified on the front panel. Upon completion of the case structure, the connection between LabVIEW and Arduino was closed. Finally, any errors encountered during this process were handled. Fig. 14 shows the front panel and Fig. 13 shows the block diagram for the LabVIEW signal transmission.

### C. Arduino-LabVIEW Interface Communication

The Arduino code was used to test whether the Arduino Uno unit could successfully execute the tasks required for monitoring muscle thickness and controlling a transducer. We wrote and uploaded sketches to the Arduino Uno, which involved reading inputs from a sensor that detects muscle thickness and outputting power to a transducer to generate vibrations. The development process included the initialization of the sensor and transducer in the setup() function and the continuous monitoring and control logic in the loop() function.

```
bool labViewInput = false;
bool deliverMedicine = false;
```

Figure 5. Two Arduino boolean variables: *labViewInput* and *deliverMedicine*. The first line flags to indicate whether the threshold has been reached. The second line flags to indicate whether to deliver the desired medicine.

Boolean variables were created to standby and wait for instructions for the threshold value and delivery commands (*Fig. 5*). In the loop() function (*Fig. 6*), data is continuously check from the serial interface. When data is available, *Serial.parseBool()* is used to read the input as a boolean value (true or false). Depending on the input value, if the input is true, the *deliverMedicine* flag is set to true. If the input is false, the *deliverMedicine* flag is set to false. This flag determines whether the transducer should be activated to deliver medicine or generate vibrations. The code then prints the received input value and the status of the *deliverMedicine* flag to the serial monitor for debugging and verification purposes.

```
void loop() {
    // Check if there's data available to read
    if (Serial.available() > 0) {
        // Read the input boolean value
        bool inputValue = Serial.parseBool();

        // Set the deliverMedicine flag based
        deliverMedicine = inputValue;

        // Print the received input value and
        Serial.print("Received input: ");
        Serial.println(inputValue);
        Serial.print("Deliver Medicine: ");
        Serial.println(deliverMedicine);
    }
}
```

Figure 6. Arduino delivery mechanism code loop

### D. Finite Element Analysis

COMSOL Multiphysics (*COMSOL Inc.*, Stockholm, Sweden) was used to determine the acoustic pressure waves and inlet flow rate velocity when the device is working. The 2D macromodel geometry was imported into the program and used as the domain to conduct the physical analysis. Numerous iterations of differing variables were checked to find the optimal values which utilized the frequency domain study for acoustic pressure and the stationary parametric sweep study for inlet flow rate. Both simulations used the velocity equation: $\sqrt{2gh}$. A height variable was arbitrarily given to calculate the parametric sweeps. For simplicity, the material of the empty space had the same properties as water. The wall of the device was determined to have a no slip boundary condition.

In the acoustic pressure simulation, the pressure speed was 40 m/s after calculating the velocity from the frequency, 40 kHz, and the distance from peak to peak, 1 mm. The angular frequency was 5000 rad/s, and the wave number was 125 rad/m as calculated from the angular frequency divided by the pressure speed. To output a pressure varied in the frequency domain, a pressure variable was created with the expression in Pascals:

$$0.02 \sin(5000t - 10x) \quad (3)$$

The turbulent flow, $\kappa \sim \varepsilon$, physics interface was used to model flow rate. With two inlets on the right side with initial values of the drug delivery velocity and one outlet on the left side with static pressure, the flow rate profile was created.

### III. RESULTS

#### A. Pressure Waves and Inlet Flow Rate Velocity

The optimal height was determined to be 2 inches above the inlet entrance with a velocity of 1.73 m/s according to the velocity equation. The transducer propagates sound waves to control the drug delivery flow rate (*Fig. 7*). The constructive waves create large signals that also cause the device to hum during activation. A slow inlet velocity fills the chamber and is stopped by a silicone duckbill valve that only opens once a pressure is felt. The pressure increases as the velocity decreases towards the center (*Fig. 8*), and as the velocity increases towards the outlet the pressure drops off. The swirling streamlines represent the fluid filling under the transducer waiting to be pushed outwards (*Fig. 9*).

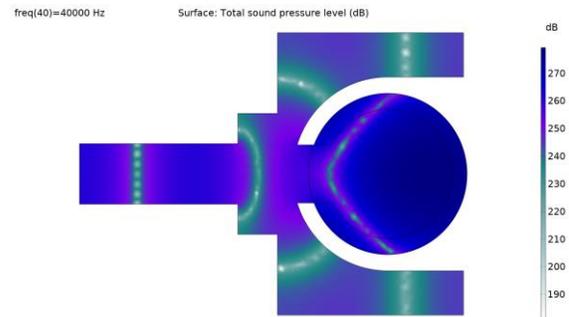

Figure 7. Acoustic Pressure Waves (dB) propagated from the center at 40 kHz.

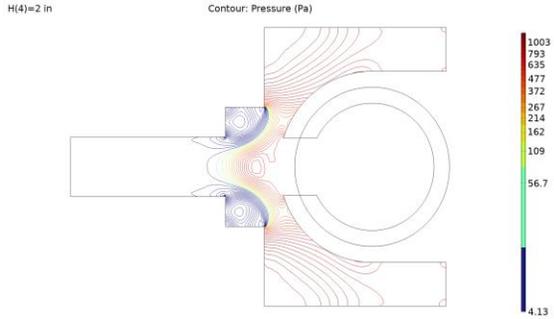

Figure 8. Fluid flow pressure (Pa) in the device through the channels.

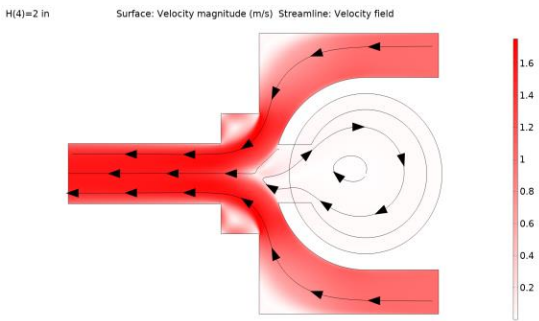

Figure 9. Velocity Magnitude of Drug Delivery (m/s) through the channels and is collected under the transducer waiting to be pushed outwards.

### B. LabVIEW VI for Muscle Biosensing

Figs. 10 – 16 display the LabVIEW block diagram along with the front panel of the virtual instrument (VI) setup presenting the data results.

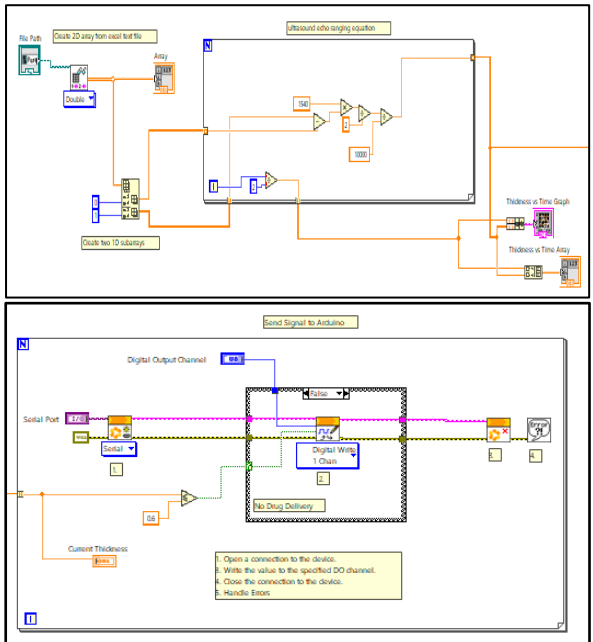

Figure 10. Full LabVIEW block diagram split before the Arduino microcontroller.

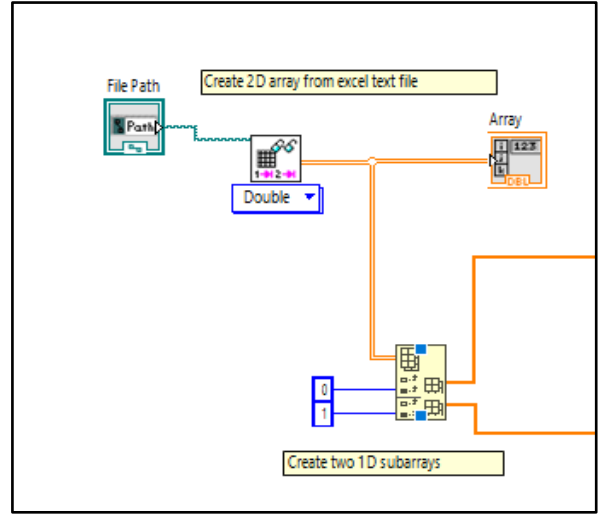

Figure 11. Raw Data Input and Array Conversion

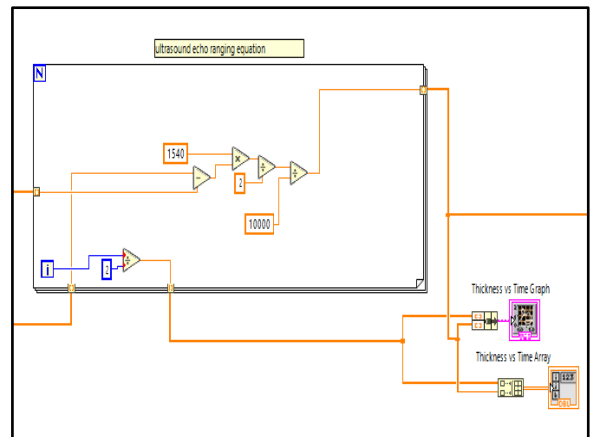

Figure 12. Ultrasound Echo Ranging Equation

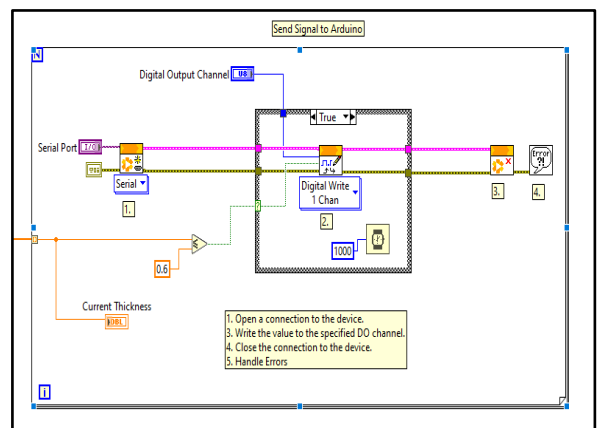

Figure 13. Signaling the Arduino Microcontroller

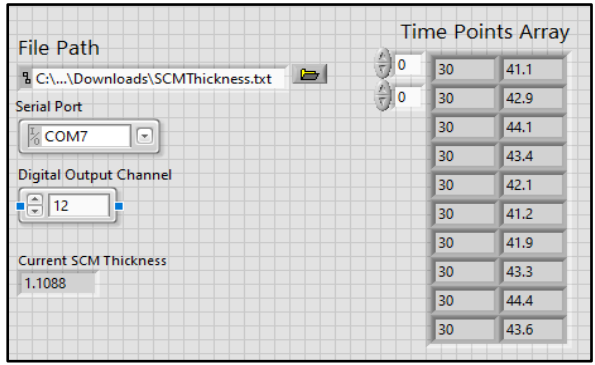

Figure 14. Raw Data Path and Time Points Array

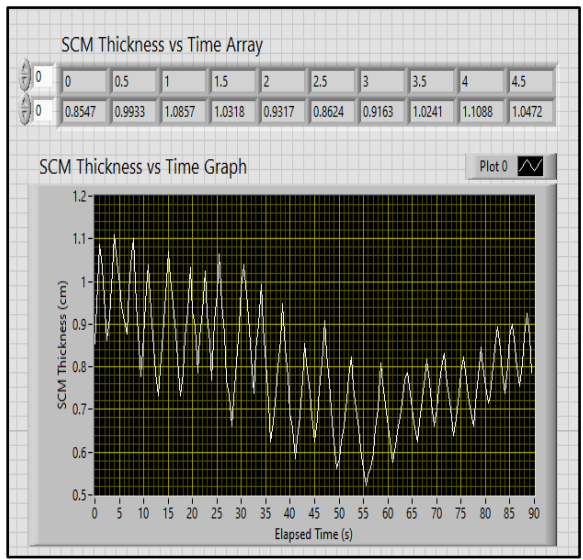

Figure 15. LabVIEW SCM thickness vs time graph created from array

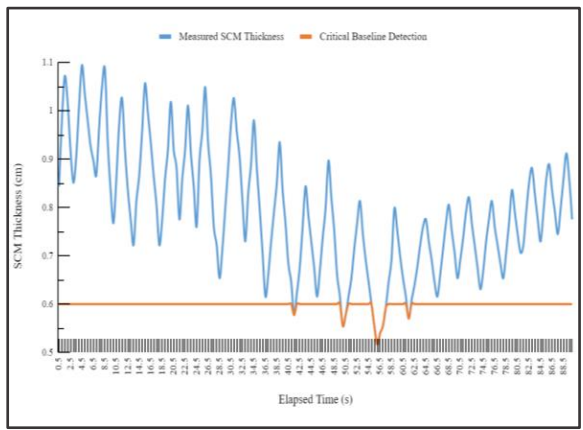

Figure 16. SCM thickness vs time graph plotted with the threshold shown

*C. Output Code Analysis*

Executing the program confirmed that the *deliverMedicine* flag accurately reflected the received input, changing its state based on the boolean values provided. The status of the *deliverMedicine* flag was printed correctly to the serial monitor, allowing the verification that the system responded appropriately to the input signals, indicating that the logic for controlling the transducer based on the input was functioning as designed. By methodically testing each task and ensuring the reliable performance of both the software and hardware components, we confirmed that our Arduino Uno was fully capable of executing the project's objectives. The successful operation of the code indicated that the Arduino platform is well-suited for this application, and we are confident in its ability to support the ongoing development of our project.

IV. DISCUSSION

To test the capabilities of the COPD device, simulations under different conditions were generated in COMSOL Multiphysics. It is assumed that with this design there will be constructive and destructive interference against the walls, but it is observed that the pressure waves from the transducer still propagate towards the outlet. The flow of liquid from the two inlets was observed at varying heights to find the optimal height to release the drug and mixing agent at. It needed to be at the height where there are no dead flow zones but also where the current is not too fast.

Once all data was collected, the team began to work on the LabVIEW code to simulate the biosensor that will be engineered to read the thickness of SCM muscles in the neck and send a signal to another device once a threshold is met. It utilized an echo ranging ultrasonic equation that calculated the distance the wave traveled in time and the speed it traveled through the muscle tissue. LabVIEW then performed mathematical calculations to output a binary output signal to Arduino to either turn on or stay idle.

The device was designed using basic materials such as silicon tubing and a silicone duckbill valve, as well as acrylonitrile butadiene styrene (ABS), the filament used to 3D print the prototype. After printing the device, it was surprisingly light even at its macroscale level when it was connected to the Arduino and laptop for testing its ability to sense and respond to changes in the input graph. Upon transitioning to the microscale, the LabVIEW code will be replaced by a physical biosensor that utilizes ultrasound to perform the same task. The physical biosensor will measure both the thickness and resistance of the muscle tissue (approximately 100 times per second). The thickness of the SCM muscle changes based on the patient's age, gender, duration of disease, and other factors. Once the thickness and impedance values drop below a certain threshold, the device will interpret the patient as experiencing acute exacerbations. Both the ultrasonic sensor and impedance recording device will be incorporated into the IC chip on the patch to minimize the size and increase power efficiency.

Once the Arduino activates the device, the muscle relaxant drug will be delivered at a constant, zero order release via microneedles into the muscle tissue or blood stream. The patch would be coded with a cooldown timer and be able to administer multiple rounds of drug to the patient before being removed. Each day, a new patch will be adhered to the neck, ready to be activated for therapeutic effect. In vitro testing will be conducted to test the microfluidic medical patch for safety and efficacy according to the ISO 10993-5 biocompatibility standards. An IRB will be established to initiate the further development and approval of the device by the FDA according to the standard approval process. Funding will be sought after to help fund all the registration and experimental requirements to meet and present the results at the end of the process. This microfluidic bioelectronic device will be submitted for an FDA approved Class II study based on 21 U.S. Code § 360.

## V. CONCLUSIONS

This medical device is in its preliminary stages since the goal is to create a microelectronic biosensing patch that attaches to the side of the neck of a COPD patient who experiences frequent acute exacerbations and has the habit of forgetting their inhaler. The plan is to use FDA-approved materials used in wearable devices that will not cause skin irritation and small battery sources for power supply.

The main future improvement is to miniaturize the device to become an adhesive patch on the neck above the SCM muscle. The outlet flow port will be designed as a diffusion-mediated microneedle array instead of outlet tubing with a duckbill valve. The two inlets will be 3D printed reservoirs designed to hold a desired amount. The low-frequency transducer uses too much voltage requiring a wall outlet, so to generate mechanical waves with less voltage, a higher frequency transducer will be required.